\documentclass[twocolumn,prb]{revtex4}

\usepackage{amsmath}
\usepackage{graphicx}
\usepackage{bm}

\begin{document}

\title{Sequential tunneling and inelastic cotunneling in nanoparticle arrays}
\author{T.~B.~Tran$^{1}$, I.~S.~Beloborodov$^{1,3}$, Jingshi Hu$^{1}$, X.~M.~Lin$^{4}$,
T.~F.~Rosenbaum$^{1,2}$, H.~M.~Jaeger$^{1,2}$ }
\address{$^{1}$James Franck Institute, University of Chicago, Chicago, Illinois 60637, USA\\
$^{2}$Department of Physics, University of Chicago, Chicago, Illinois 60637, USA\\
 $^{3}$Materials Science Division, Argonne National Laboratory, Argonne, Illinois 60439, USA\\ $^{4}$Center of Nanoscale Materials, Argonne National Laboratory, Argonne, Illinois 60439, USA}

\date{\today}

\begin{abstract}
We investigate transport in weakly-coupled metal nanoparticle arrays, focusing on the regime where tunneling is competing with strong single electron charging effects. This competition gives rise to an interplay between two types of charge transport. In sequential tunneling, transport is dominated by independent electron hops from a particle to its nearest neighbor along the current path.  In inelastic cotunneling, transport is dominated by cooperative, multi-electron hops that each go to the nearest neighbor but are synchronized to move charge over distances of several particles.   In order to test how the temperature-dependent cotunnel distance affects the current-voltage ($I-V$) characteristics we perform a series of systematic experiments on highly-ordered, close-packed nanoparticle arrays.  The  arrays consist of $\sim 5.5$nm diameter gold nanocrystals with tight size dispersion, spaced $\sim 1.7$nm apart by interdigitating shells of dodecanethiol ligands. We present $I-V$ data for mono-, bi-, tri- and tetralayer arrays.  For stacks 2-4 layers thick we compare in-plane measurements with data for vertical transport, perpendicular to the array plane.  Our results support a picture whereby transport  inside the Coulomb blockade regime occurs by inelastic cotunneling, while sequential tunneling takes over at large bias above the global Coulomb blockade threshold $V_t(T)$, and at high temperatures.
For the smallest measurable voltages, our data is fit well by recent predictions for the temperature dependence zero-bias conductance due to multiple cotunneling. At finite, but small bias, the cotunnel distance is  predicted to set the curvature of the nonlinear $I-V$ characteristics, in good agreement with our data.  The absence of significant magnetic field dependence  up to 10T in the measured $I-V$ characteristics further supports the picture of inelastic cotunneling events where individual electrons hop no further than the nearest neighbor. At large bias, above the global Coulomb blockade threshold, the $I-V$ characteristics follow power law behavior with temperature-independent exponent close to $2$ predicted for sequential tunneling along branching paths that optimize the overall charging energy cost.
\end{abstract}

\maketitle

\section{Introduction}

 Arrays of closely-spaced nanoparticles are currently becoming a model system for fundamental studies of mesoscopic charge transport and for targeted applications in nanotechnology.~\cite{Heinzel} Quite generally, their behavior depends on the competition of several energy scales that can be controlled independently by tuning the properties of the particles and their geometric arrangement. These are the mean energy level spacing, the Coulomb charging energy for a single particle, as well as the tunnel energy associated with the inter-particle coupling.~\cite{Beloborodov2}

Even for particles that individually are simple metals, the array as a whole can exhibit the full range from strong, exponential  ("insulating") to weak, non-exponential ("metallic") temperature dependence of the conductance.  This behavior can be tuned by the strength of coupling between the building blocks - the bare, high-temperature tunneling conductance $g$.   In the regime of strong coupling,  the Coulomb interaction is screened and electrons propagate easily.  In the opposite limit of the weak coupling,  single electron charging becomes significant, leading to Coulomb blockade behavior and the localization of electrons on individual particles.~\cite{Averin92} As a results, arrays with strong coupling $g > g_q$, where $g_q = e^2/h$ is the quantum conductance, behave as metals, while arrays with $g < g_q$ are insulators in the low-temperature, low-bias limit.

The overall, global transport properties arising from local, charging-mediated tunneling have been investigated experimentally for a number of systems formed from metallic or semiconducting nanoparticles.~\cite{Bezryadin, Yu, Parthasarathy, Parthasarathy2,Elteto, Andres, Kurdak, Black, Ancona,Zabet-Khosousi,Herrmann,Bufon} Close-packed two-dimensional arrays of highly size-controlled metal nanocrystals, in particular, have served to elucidate the effect of structural disorder, such as imperfections in the particle arrangements, on the current-voltage characteristics~\cite{Parthasarathy, Ancona} and to test predictions~\cite{Middleton} for quenched charge disorder.~\cite{Parthasarathy2, Elteto}  Charge disorder arises from variations in the local chemical potentials, due to polarization by trapped parasitic charges in the substrate or in the ligand shells surrounding the particles. In practice, charge disorder is unavoidable for arrays involving more than a few particles and, as a consequence, tunneling occurs along paths that optimize the overall energy cost. This results in two distinct regimes.  At applied bias voltages large enough to overcome local Coulomb energy costs, transport occurs by sequential tunneling between neighboring particles along a set of optimal paths.~\cite{Middleton,Elteto}  At small bias and low temperatures, sequential tunneling is suppressed by the Coulomb blockade.  In this regime, conduction involves higher-order, cooperative tunneling processes, so-called cotunneling events, that can transport charge over distances of several particles without incurring the full Coulomb energy costs.~\cite{Averin,Beloborodov2,Feigel'man,Beloborodov} By virtue of the uncertainty principle, this is possible through the creation of short-lived excitations on the intermediate particles.  In arrays of metallic nanoparticles and for temperatures from a few Kelvin on up, these excitations consist of electron-hole pairs, produced by cotunneling events in which one electron tunnels onto a particular particle while simultaneously another, lower energy electron tunnels off (inelastic cotunneling).

 Recently we investigated electron transport through large, highly-ordered metal nanoparticle arrays with lateral size of 65-70 particles between the electrodes.~\cite{Tran} We found that the resulting, current-voltage characteristics as well as the temperature dependence of the conductance for small applied bias voltages are in line with predictions based on multiple inelastic cotunneling.~\cite{Beloborodov,Beloborodov2,Feigel'man} In particular, our results indicated that low-temperature, low-bias transport across the full array occurs via a sequence of cotunneling events, each involving a few particles (up to $\sim 4$), depending on temperature.  This temperature-dependent cotunneling distance was interpreted as resulting from an optimization process that weighs the likelihood of n-electron, cooperative cotunneling events against the net energy cost.

 In the present paper we extend these results and specifically address the question about what sets the cotunneling distance.  This characteristic length scale not only determines the low-bias current-voltage characteristics, but also the temperature dependence of the zero-bias conductance. We present results from experiments in which the distance between electrodes has been tuned to become commensurate with the cotunneling distance.  These experiments involve vertical transport through carefully prepared stacks of particle layers, up to four layers in height. This approach provides a means to control the number of tunnel junctions the electron traverses to get across the electrodes, allowing us to test the cotunneling scenario in detail.

This paper is organized as follows.  In Sec.~\ref{IV} we discuss general features of the current-voltage ($I-V$) characteristics predicted for metal nanoparticle arrays. Sec.~\ref{experiment} contains details about the sample fabrication and the experimental set-up.  We present in Sec.~\ref{expresult} our experimental results for the evolution of the $I-V$ characteristics as a function of temperature and stack thickness. We compare results obtained from vertical transport through stacks of layers to in-plane measurements on monolayers and tetralayers and discuss these data within the context of recent models for charging-energy-mediated tunneling transport. Sec.~\ref{concl} contains brief conclusions.

\section{Current-Voltage characteristics: Background}
\label{IV}

The purpose of this section is to discuss the general features of the current-voltage characteristics predicted for nanoparticle arrays in  the semi-classical Coulomb blockade regime and in the cotunneling regime. The semiclassical or "orthodox theory" of Coulomb blockade predicts exponential suppression of the conductance at low temperatures.~\cite{Glazman} In the zero-bias limit this suppression leads to $g_0(T) = g\exp [-E_C/k_B T]$ for  transport through a single nanoparticle connected by tunnel junctions to electrodes.  Here $E_c$ is the charging energy, for a weakly coupled particle approximated by $E_C = \frac {e^2}{4\pi\epsilon\epsilon_0 a}$, where $\epsilon$ is the dielectric constant, $\epsilon_0$ the permittivity of free space, and $a$ is the particle radius.  The bare conductance $g$ corresponds to the temperature-independent Ohmic tunnel conductance a junction would have in the absence of all charging effects. For nanoparticles, small $a$ implies large $E_C$ and thus a significant suppression of sequential tunneling at low temperatures.  Conduction then occurs by one of two means:  At sufficiently large applied bias voltage, the Coulomb blockade is overcome so that sequential tunneling can commence.  In arrays, this happens for $V > V_{t}$, where $V_{t}$ is a global Coulomb threshold voltage discussed below.  Conversely, while a Coulomb blockade prevents direct sequential tunneling  for $V < V_{t}$, cotunneling becomes a viable conduction channel.  We note here that, for logic or switching devices based on single electron tunneling, cotunneling typically needs to be avoided. In principle, arrays are well-suited for this purpose since the probability for system-spanning cotunnel events decreases dramatically with increasing system array size while $V_{t}$ increases linearly.~\cite{Middleton,Elteto}  Nevertheless, as we show here, multiple shorter cotunnel events still can contribute significantly to the overall conduction.

\subsection{Semi-classical regime, $V > V_{\mathrm{t}}$}

Applying the semiclassical picture to large arrays of nanoparticles with a random distribution of local chemical potentials, Middleton and Wingreen showed that, at $T = 0$K, conduction only occurs beyond a global threshold voltage, $V_t= \alpha NE_C$, where N is the number of nanoparticles spanning the gap between the electrodes, and $\alpha$ is a prefactor around 0.2-0.5 that depends on dimensionality and array geometry.~\cite{Middleton} At $T=0$, in this model there is no conductance for $V<V_t$. Above $V_t$, the electrons percolate through the array via a multitude of branching paths that navigate local Coulomb blockade thresholds and optimize the total charging energy cost. Non-linear $I-V$ characteristics with a power law dependence on $V-V_t$ emerge as a consequence of the branching.~\cite{Middleton}

The semiclassical $T=0$ scenario for transport above a global Coulomb threshold can be extended to finite temperatures.~\cite{Parthasarathy} As long as the charge disorder is sufficiently strong (an assumption easily satisfied in nanoparticle arrays because $E_C$ is large), the main effect of temperature is to  wash out differences between chemical potentials of neighboring particles and thus  "erase" local Coulomb blockade thresholds.  As a consequence, the global threshold $V_t$ decreases linearly with temperature, while the power law exponent in the $I-V$ characteristics remains unaffected.

This picture is supported by experimental results on a variety of metal nanoparticle systems.~\cite{Ancona,Bezryadin,Cordan} Specifically, for two-dimensional (2D) close-packed Au particle arrays it was found that $V_t(T) = V_t (1-T/T^*)$ where $V_t= \alpha NE_C$ with $\alpha=0.23$ and a temperature-independent exponent close to 2.2.~\cite{Parthasarathy2,Elteto} The weak, linear dependence of the global threshold on temperature  is very robust and persists in the presence of structural disorder in the array.  On the other hand, such disorder leads to $I-V$ characteristics that deviate from a simple power law above $V_t(T)$.~\cite{Parthasarathy,Blunt}

By design, the Middleton and Wingren model and its extensions only apply to the large bias limit and do not consider thermally activated charge excitations.  The low-bias conductances across "non-erased" local thresholds are assumed to be exponentially suppressed and, as a result, the global low-bias conductance is effectively controlled by the largest threshold along the optimal path.  Because this path can avoid high charging energy costs to some extent by branching, for 2D close-packed arrays the relevant activation energy is not the full charging energy, but instead $U \approx 0.2 E_C$ and which leads to a zero-bias conductance $g_0(T) = g\exp [-U/k_B T]$.~\cite{Parthasarathy,Elteto}  This simplified picture neglects processes that involve charge transfer over distances larger than a single particle (such as the cotunneling processes discussed next),  but it is applicable at sufficiently high temperatures where the zero-bias conductance crosses over to simple activated, Arrhenius behavior characteristic of nearest neighbor hopping.

\subsection{Cotunneling regime, $V<V_{\mathrm{t}}$}

Cotunneling, first discussed by Averin and Nazarov, is a tunneling process whereby electronic charge is transferred through several neighbor particles cooperatively~\cite{Averin}. For sequential tunneling each electron hop  is a separate quantum event. By contrast, in cotunneling all charge transfers from the initial particle to the $n^{th}$ final particle occur via virtual states and thus count as a single quantum event~\cite{Glazman}.  In principle, there are two different types of cotunneling processes: elastic and inelastic. In elastic cotunneling, an electron of the same eigenstate tunnels from the initial to the final particle, while inelastic cotunneling involves the cooperative motion of {\it multiple} electrons, tunneling from one particle to the nearest neighbor in concert. Because elastic cotunneling becomes significant only at very low temperatures (below $\approx$ 1K in our system) we will focus in the following on inelastic processes.

 For the low-bias regime, the key quantity to consider for calculating the conductance is the probability to create an electron-hole excitation from an initially neutral particle.  This probability is proportional to $ \exp [-E/ k_B T]$, where $E$ is the electro-static energy associated with the electron-hole pair.  If charge transport occurs by sequential tunneling, the electron-hole pair is separated by one particle after the event and $E$ is simply $E_C$.   For cotunneling the final separation can be larger.  If, at the end of an inelastic cotunneling process along a chain of neighboring particles, $j$ electrons have each moved one particle down the current path, the electron-hole pair is separated by $j+1$ particles, and $E =  \frac{E_C}{j}$.   As a result, for $k_BT << E_C$ cotunneling will always give rise to a larger zero-bias conductance than sequential tunneling (even though cotunneling is a higher order process whose amplitude scales as $(g/g_q)^j$, see below)~\cite{Beloborodov, Beloborodov2}.

 For the limit of zero applied bias, it was shown~\cite{Beloborodov, Beloborodov2} that cotunneling events span a typical distance, $r^*$. In terms of the number of tunnel junctions involved, this distance is given by $r^*/d= (\frac {E_C}{k_B T}\,
 \frac{\xi a}{d^2})^{1/2}$, where $\xi$ is the localization length ($\xi \leq 2 a$ for weakly coupled particles) and $d$ the particle center-to-center spacing ($d \approx 8nm$ in our arrays). Equating this distance with $j$ in the formula above for the activated creation of a dipole excitation leads to
\begin{subequations}
\begin{equation}
\label{eq:G0}
g_0(T) =  g \exp(-\sqrt{T_0/T}),
\end{equation}
for the zero-bias conductance, with a characteristic temperature $T_0$ given by
\begin{equation}
\label{eq:T0}
T_0 = \frac{C E_C a}{k_B \xi}=\frac {C e^2}{4 \pi \epsilon \epsilon_0 k_B \xi},
\end{equation}
\end{subequations}
where $C \simeq 2.8$.~\cite{Shklovskii} This result has the same functional form for the temperature dependence as Efros-Shklovskii variable range hopping (VRH) for doped semiconductors.~\cite{Shklovskii} It similarly describes the effect of balancing net tunnel distance and energy cost. However, it arises here not from direct, single charge tunnel events between sparsely distributed atomic defect sites, but from cooperative multi-electron processes in dense arrays of metal particles.

At finite but small voltage, for inelastic cotunneling processes involving a distance of $j$ junctions, the current is given by
\begin{equation}
\label{eq:In_current}
I  \sim V_{\mathrm{jct}}\left[\frac{g}{g_q}\right]^j e^{\frac {jeV_{\mathrm{jct}}-E}{k_BT}} \left[\frac{(eV_{\mathrm{jct}})^2+(k_BT)^2}{E_C^2}\right]^{j-1}.
\end{equation}
Here $V_{\mathrm{jct}}$ is the bias voltage drop across each of the $j$ junctions, $g$ and $g_q$ are the junction and quantum conductances respectively, and $E =\frac{E_C}{j}$ the energy cost for the dipole creation.  For $V_{\mathrm{jct}} << k_BT$ the conduction is Ohmic and optimization with respect to $j$ returns the result for $g_0(T)$ in Eq.~(\ref{eq:G0}).

If the current path involves sufficiently many particles in series, we expect that multiple cotunnel events will occur, each over a typical distance $j=r^*/d$. This distance will increase according to $T^{-1/2}$ as the temperature is reduced,  leading to a corresponding increase in the curvature of the low-bias $I-V$ characteristics.  On the other hand, if the length of the current path is fixed at a distance shorter than $r^*/d$, the temperature dependence of the curvature will saturate. This paper explores these effects by performing transport measurements in metal nanoparticle arrays of different thicknesses in the direction perpendicular to the array plane. We compare these results with those of monolayers and tetralayers where current-voltage measurements were taken in-plane.

\section{Experimental procedure}
\label{experiment}

\begin{figure}
\includegraphics[width=3in]{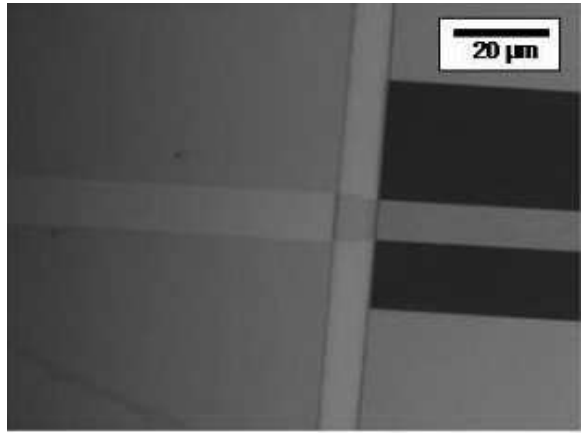}
\caption{Optical image of a bilayer nanoparticle array on a  $Si_3N_4$ window substrate. The top and bottom electrodes (light rectangular strips) are oriented perpendicular to one another. The large dark rectangle is the window area.\label{fig:optical}}
\end{figure}

Dodecanethiol-ligated gold nanoparticles were synthesized by the digestive ripening method described in Ref.~\cite{Lin}. This provided particle diameters around 5.5nm with tight size control and dispersion less than $5\%$ in each batch. The nanoparticle arrays were deposited onto silicon substrates coated with 100nm amorphous $Si_3N_4$. For sample characterization by transmission electron microscopy (TEM), these substrates contained $70 \mu$m $\times 70 \mu$m or $300 \mu$m $\times 300 \mu$m "window" areas under which the Si had been etched away to leave free-standing, TEM-transparent $Si_3N_4$ membranes. Before particle deposition, we created by electron beam lithography and thermal evaporation 20nm thick chromium electrodes on top of the substrates, reaching into the window areas. For in-plane measurements, sets of two opposing electrodes were fabricated, $2 \mu$m in width and with gap $500$nm between them.  For transport measurements perpendicular to the array plane, the bottom electrode consisted of a Cr strip $10 \mu$m in width and $600 \mu$m in length. After particle deposition, a top electrode was evaporated over the array, consisting of a Ni strip $30$nm in thickness, 300$\mathrm{\mu m}$ in length and $10 \mu$m in width. The top electrodes  were shadow evaporated through masks made from the same $Si_3N_4$ window substrates used for the samples, with $CF_4$ etched rectangular holes of dimension $300 \mu$m $\times 10 \mu$m.  By orienting the top electrodes  orthogonal to the bottom electrodes in a cross-like geometry, overlapping regions of area $10 \mu$m $\times 10 \mu$m were created.  Cr and Ni were chosen as the electrode material because of their excellent adhesion to the  $Si_3N_4$ as well as to the gold particles, and because uniform, continuous electrodes can be formed reliably at relatively low thickness.  The use of Ni kept the heat load on the array during evaporation lower than would have been the case with Cr.  However, the uppermost dodecanethiol ligands were likely burnt away during this process, bringing the top electrode in direct contact with top particle layer.  This is substantiated by the vertical transport measurements discussed below.  On the other hand, we were able to ascertain by TEM that the top electrode evaporation did not lead to any noticeable sintering of the particles.  Figure~\ref{fig:optical} displays an optical image of a bilayer sample with top and bottom electrodes.

\begin{figure}
\includegraphics[width=3in]{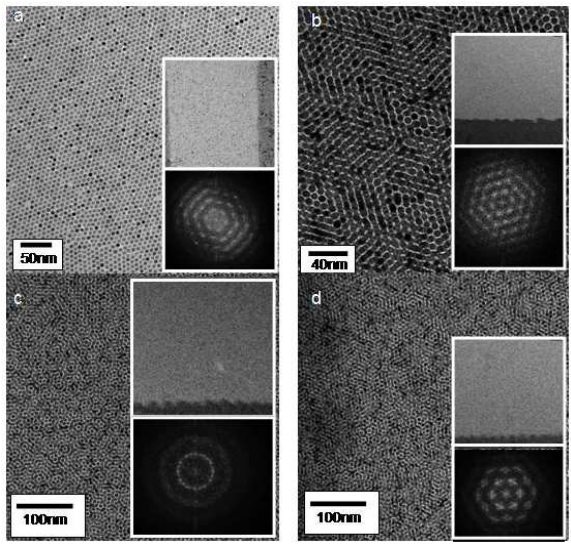}
\caption{Transmission electron microscopy (TEM) images of nanoparticle arrays used in the experiments: (a) monolayer, (b) bilayer, (c) trilayer, and (d) tetralayer. The main panels show close-up views of the highly ordered particle arrangement. The bottom insets in a-d display diffraction patterns computed by fast Fourier transform. The top insets  are zoomed-out images of the area covering the region between the in-plane electrodes of the monolayer (a), and the region near the bottom electrodes of bi-, tri-, and tetra-layers measured in perpendicular direction (b-d). The dark regions of in these insets are the electrodes.
\label{fig:im_verttransport}}
\end{figure}

Gold nanoparticle monolayers were created at the air interface of a water droplet, similar to a Langmuir-Schaefer technique~\cite{Roberts}. First, a $\sim 5$mm diameter water droplet was created on a glass slide. This was followed by deposition of 20-30$\mu$L of gold nanoparticle solution (concentration $10^{13}$ particles/mL) on top of the droplet. Using this technique, the particles quickly spread to form a beautiful, compact monolayer across the water surface, presumably due to surface tension. By dipping an inverted substrate with prefabricated electrodes onto the top of the droplet, the monolayer was then transfer-printed onto the substrate. This preparation method gave interparticle distances of $1.7$nm to $1.8$nm between different batches of nanoparticle solutions. Multilayer arrays were created via layer-by-layer deposition, repeating the process described above. There is no registry between subsequent layers using this technique, but the compact, highly ordered particle arrangement is maintained. Figure~\ref{fig:im_verttransport}a shows a monolayer spanning the region between two planar electrodes. TEM images of  multilayer arrays for vertical transport measurements in Figs.~\ref{fig:im_verttransport}b-d were taken right next to the electrodes because the top electrodes in these samples were not TEM-transparent. The data presented in this paper was taken on mono-, bi-, tri, and tetralayers.

Current-voltage ($I-V$) measurements were performed with a Keithley 6430 source meter. Bias voltages of $V\le \lvert 4V \rvert$ were applied perpendicular to the sample plane at rates of 0.05-1.5mV/sec and at temperatures 5K-100K. For in-plane $I-V$ measurements, bias voltages $V<\lvert 20V \rvert$ were applied at rates of 5-25mV/sec over the range from $10$K - $160$K.  The cryostat was wired with coax cables for low-leakage, low-noise measurements.  Typical current noise was below 40fA. TEM images of all samples were taken posterior to $I-V$ measurements to preclude any irradiation damage or contamination of the samples, even if undetectable in the image, from affecting the transport measurements.

The zero-bias conductance was obtained by performing $I-V$ measurements, at fixed temperature, covering a narrow voltage range across the origin and extracting $g_0(T)$ from the slope. Care was taken to restrict this voltage range to the linear response.  Because of the current noise floor, this meant that $g_0(T)$ could be tracked reliably down to temperatures around 20K, but not always below that value.  This same issue also restricted the extent over which we could track the nonlinear behavior at the lowest temperatures before becoming limited by the noise floor.

Transport measurements in the presence of the magnetic field were performed in a $14/16$T Oxford He-3 cryostat. Magnetic fields up to 10T were applied both parallel and transverse to the sample plane at temperatures of $5$K, $1$K and $400$mK. The voltage sweep rate was $0.5-1.2$ mV/s.

\section{Results and Discussion}
\label{expresult}
\subsection{Non-Ohmic Regime}
We first focus on the regime of finite applied bias. Here the transport properties are highly non-Ohmic. Figs.~\ref{fig:IVs_verticaltransport} and~\ref{fig:IVs_inplane} show our $I-V$ data for vertical and in-plane transport, respectively.  In both figures, the left and right columns show the same data, but on linear and on double logarithmic scales. By appropriately normalizing the current so that $I-V$ traces taken at different temperatures can be visualized on the same linear plot, the increase in nonlinearity with decreasing temperature is immediately apparent.  This is the purpose of the left columns of panels.  The right columns show the raw data and indicate how the highly nonlinear and strongly temperature-dependent behavior at finite, but small, bias merges into a much less temperature-dependent high bias regime.  For in-plane transport across large arrays this high bias regime corresponds to the semiclassical $V>V_t$ regime.  Here, all $I-V$ traces are predicted to  approach the same power law behavior with exponent close to 2, merging onto this asymptote at different temperatures given by $V_t(T)$.  This is borne out by the data in Fig. \ref{fig:IVs_inplane}c,d.  For vertical transport, the distance traversed at high bias is the thickness of the stack of layers (a few particle distances), and this is too short for a path branching pattern to emerge. Instead, the behavior seen in Fig.~\ref{fig:IVs_verticaltransport} resembles that of short parallel paths, each retaining the strongly temperature-dependent prefactor in Eq.\ref{eq:In_current} (as opposed to acquiring the weak, linear T-dependence of $V_t(T)$).

\subsubsection{Vertical Transport}

\begin{figure}
\includegraphics [width=4in]{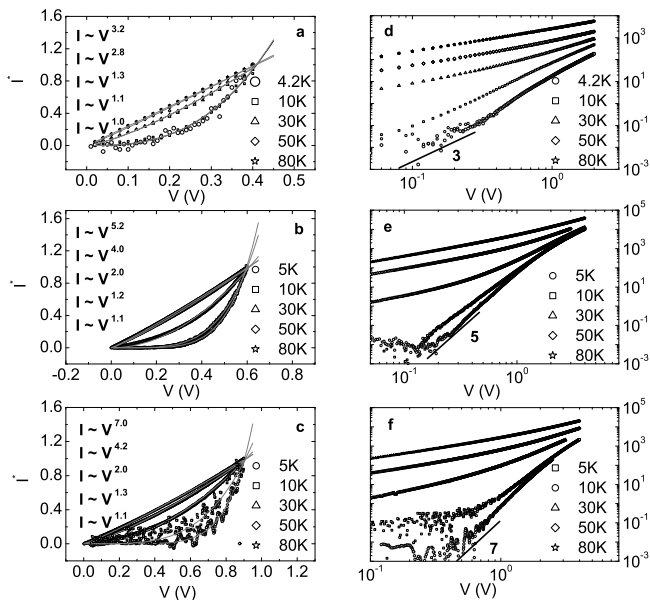}
\caption{Current-voltage ($I-V$) characteristics as a function of temperature for bilayer (a), trilayer (b),  and tetralayer (c) arrays measured in the direction perpendicular to the array plane. Panels d, e and f are log-log plots of the data in panels a, b and c, respectively; the solid lines are guides to the eye corresponding to power laws with exponents as indicated. 
The current scales in panels a-c are normalized with respect to the current at fixed bias voltage (0.4V in (a), 0.6V in (b), 0.9V in (c)) to bring out the increasing nonlinearity as temperature is reduced. The solid lines in panels a-c correspond to power law fitting curves of the form $I \sim V^{\beta}$. The exponents $\beta$ obtained from these fits have uncertainties of $\pm 0.1 \mathrm{to} \pm 0.2$ in panels a and b, and $\pm 0.3$ in panel c. \label{fig:IVs_verticaltransport}}
\end{figure}

As can be seen from data on bi-, tri- and tetra-layers (Fig.~\ref{fig:IVs_verticaltransport}a-c), the curvature of the small bias, low-temperature $I-V$ characteristics increases with increasing thickness of the stack.  The data are fit well by power laws of the form $I \sim V^{\beta}$, with exponents $\beta$ as indicated in the figure. At 5K, the exponents are approximately 3, 5, and 7 for a bi-, tri- and tetralayer, respectively. While $\beta$ for the bilayer remains the same as the temperature increases to 10K, $\beta$ for tri- and tetralayers decreases to 4 and 5.

In order to compare these data with theoretical predictions, we first estimate the charging energy. Using $E_C = \frac{e^2}{4\pi \epsilon \epsilon_0 a}$, particle radius $a  \approx 2.5$nm as measured by TEM and $\epsilon \approx 4$ we find $E_C \approx 125\mathrm{meV}$ or $1600\mathrm{K}$.  Here the value of the dielectric constant is based on our previous results on multilayers.~\cite{Tran}

For vertical transport across a short stack of $n$ particles, the applied bias voltage and the bias per junction are related by $V_\mathrm{jct} = \frac {V}{n}$.  Based on Eq.~(\ref{eq:In_current}) we therefore expect that the nonlinear cotunneling regime $ k_B T < e V_\mathrm{jct} < E_C$ should persist up to a good fraction of a Volt applied bias at low temperatures.  This is indeed what we find in Fig.~\ref{fig:IVs_verticaltransport}.  Equating the exponents $\beta$ found from the power law fits to the exponent $2j-1$ in Eq.~(\ref{eq:In_current}) we find the typical number of junctions participating in the cotunnel events. At the lowest temperatures, this leads to $2$, $3$ and $4$ junctions for bi-, tri- and tetralayers, respectively, consistent with the notion that the top electrode has fused directly to the uppermost particle layer. The fact that $\beta$ for the bilayer saturates at a value around 3 below 10K is an indication that the stack thickness provides a cut-off: dipole separations larger than 2 junctions are not possible because they would exceed the distance between the top and bottom electrodes (however, see the caveat for the zero-bias limit, below). Consequently, at these low temperatures all cotunneling events span the electrode distance, the exponential term accounting for the thermal creation of the electron-hole pair inside the array no longer remains, and Eq.~(\ref{eq:In_current}) reduces to the the form known from short linear arrays of mesoscopic islands~\cite{Averin,Tran}
\begin{equation}
\label{eq:I_inelasticper}
I_\mathrm{in} \sim \frac{V}{j} \left[\frac{g}{g_q}\right]^j \left[\frac{eV}{j\, E_C}\right]^{2j-2} [ 1+j^2(j-1)(\frac{k_BT}{eV})^2],
\end{equation}
where the term in the last square brackets was obtained by expanding $[1+(\frac{k_BT}{eV})^2]^{j-1}$. As before, the power-law exponent equals $2j-1$ in the voltage range $k_BT < \frac{eV}{j} < E_C$ and reduces to 1 at high temperatures.  Using Eq.~\ref{eq:I_inelasticper}, appropriate for small $\frac {k_B T}{eV}$, we find that the current between 4.2K and 10K is predicted to increase by a factor of 6. This is in reasonable agreement with the data which shows a factor of 8-10. For the tri- and the tetralayers, the cotunneling length is on the order of the sample length only at 5K. At higher temperatures, this length becomes less than the distance between the electrodes, and thus multiple cotunneling will be involved.

\subsubsection{In-plane Transport}

\begin{figure}
\includegraphics[width=3.5in]{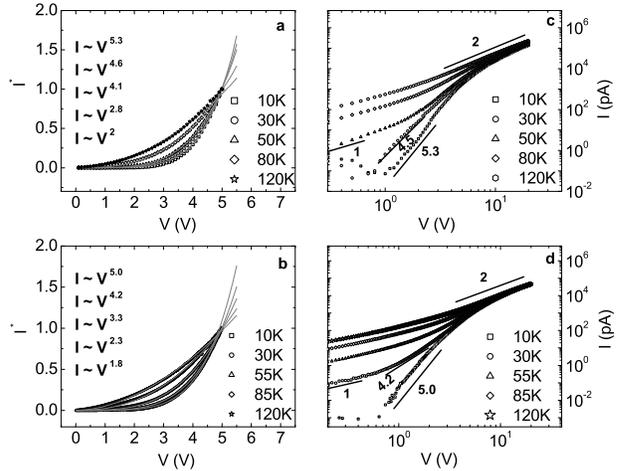}
\caption{Current-voltage ($I-V$) characteristics as a function of temperature for monolayer (a) and tetralayer (b) arrays measured along the array plane. Panels c, d are log-log plots of the data in panels a, b, respectively. The current scales in panels a and b are normalized with respect to the current at fixed bias voltage of 5V to bring out the increasing nonlinearity as temperature is reduced. The solid lines in panels a, b correspond to power law fitting curves of the form $I \sim V^{\beta}$. The exponents $\beta$ obtained from these fits have uncertainties of $\pm$ 0.2 to $\pm$0.3. \label{fig:IVs_inplane}}
\end{figure}

$I-V$ measurements taken in-plane for comparison exhibit behavior similar to the perpendicular measurements just discussed. This is shown in Figs.~\ref{fig:IVs_inplane} for a mono- and a tetralayer (for additional data see Ref.~\cite{Tran}). Because these are large arrays where the number of junctions across the gap is $\sim 70$, the global Coulomb blockade threshold $V_t \sim NE_C$ easily reaches several Volts.  As for vertical transport, the data in the small-bias regime below $V_t$ are fit well by power laws of the form $I \sim V^{\beta}$, with exponents $\beta$ as indicated in the figure.   Equating $\beta$ with the exponent $2j-1$ in Eq.~(\ref{eq:In_current}) similarly gives the typical cotunnel distance $j$ as a function of temperature. The  power-law exponents for both the monolayers and tetralayers measured in-plane imply that the maximum number of particles participating in one cotunneling event is much smaller than the number of particles spanning the gap between the electrodes, thus necessitating multiple cotunnel events to get across. The fact that the tetralayer's power-law exponent $\beta \sim 5$ at $10K$ is similar to the value for the same exponent measured at 10K perpendicular to the array plane ($\sim 4.2$), supports the notion that the same cotunnel mechanism is at work leading to a similar typical cotunnel distance in both cases. For $V>V_t$, all $I-V$ curves turn over to approach a power-law exponent close to $2$ as predicted by the semiclassical picture, and for $V$ approaching the zero-bias limit, the $I-V$ traces become linear, in line with Eq.~(\ref{eq:In_current}).

\subsection{Ohmic Regime}

Similar to our observations on samples measured in-plane~\cite{Tran},  $g_0(T)$ in the perpendicular direction also follows the stretched exponential form given by Eqs.~(\ref{eq:G0}) and~(\ref{eq:T0}).  As mentioned earlier, this behavior echoes VRH, but here the same functional form with the characteristic inverse square root temperature dependence in the exponential arises from multiple cotunneling. Interestingly, we find this functional form for all stack thicknesses.  In interpreting this result, one potential problem is that, for the vertical transport case discussed here, a straight path form the bottom to the top electrode involves only a few junctions (no more than perhaps 4 for the tetralayer). Therefore, at first glance {\it multiple} cotunnel events and the optimization of cotunnel distance versus energy cost, both necessary to produce the inverse square root temperature dependence of $g_0(T)$, appear unlikely. However, it is reasonable  that the paths taken by the electrons depend on the presence of the applied field. As $V\rightarrow 0$, the electrons might be able to meander into the lateral, in-plane direction to find an optimum, energetically cost-effective path to tunnel from the bottom to the top electrode. At higher applied bias, the electric field channels the electrons along a more direct path, and thus the cotunnel distance cannot exceed the number of tunnel junctions across the stack thickness. We speculate that this scenario is the reason why the low-temperature $I-V$ characteristics, measured at finite bias, indicate a relatively small number of junctions commensurate with the stack thickness, while the zero-bias conductance, measured at a significantly smaller bias (feasible at higher temperatures), hints at much longer paths.

\begin{figure}
\includegraphics[width=3.5in]{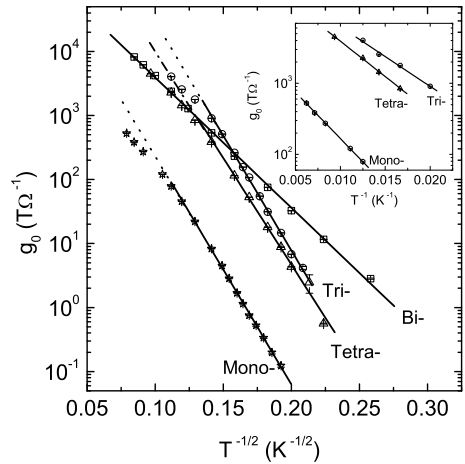}
\caption{Temperature-dependent zero-bias conductance of a monolayer measured in-plane, and of a bilayer, trilayer and tetralayer where $I-V$ measurements were orthogonal to the sample plane. The inset shows Arrhenius behavior in the monolayer with $I-V$ measurements performed in the lateral direction, and in the trilayer and tetralayer with $I-V$ measurements taken perpendicular to the sample plane. \label{fig:g0T}}
\end{figure}

From the slopes of the semi-log plots in Fig.~\ref{fig:g0T}, we find the characteristic temperatures, $T_0$ and, using Eq.~\ref{eq:T0}, the localization lengths, $\xi$ for bi-, tri- and tetra-layers.  Table~\ref{table:T_0} compares $T_0$ values from both in-plane and perpendicular measurements (including data from Ref.~\onlinecite{Tran}). All values are between 72-83K, except  for vertical transport through a bilayer.  The $T_0$ in this case is significantly smaller, implying correspondingly larger values for either the localization length or the dielectric constant (see Eq.~\ref{eq:T0}).   Since the localization lengths obtained from $T_0$  for all other systems are $\approx 2$nm, and thus less than a particle diameter as appropriate for cotunneling through weakly coupled particles, it seems reasonable that the reduced $T_0$ is associated with an increase in the effective $\epsilon$.  This is likely to be produced by the close proximity of the top electrode required for vertical transport. As soon as the electrodes are 1 to 2 layers away, the dielectric constant of the systems seems to be unaffected, i.e., the $T_0$ of both trilayers and tetralayers are similar. The same argument might also explain why the in-plane monolayer exhibits a slightly larger $T_0$ than most other systems: it experiences less screening from additional layers above it.

\begin{table}[h]
\centering
\begin{tabular}[t]{|c|c|c|}
\hline
Sample & $T_0(K)$ (in-plane) & $T_0(K)$ (Perpendicular) \\
\hline
Monolayer &$6.5 \times 10^3$ &  ----- \\
\hline
Bilayer &$3.3 \times 10^3$  & $2 \times 10^3$ \\
\hline
Trilayer &$5.1 \times 10^3$ & $6.4 \times 10^3$ \\
\hline
Tetralayer &$5.2 \times 10^3$ & $5.8 \times 10^3$ \\
\hline
\end{tabular}
\caption{Comparison of the characteristic temperatures $T_0$ obtained from fits of the zero-bias conductance data to Eq.~\ref{eq:T0}. $T_0$ values have uncertainties of $\sim 5\%$.}
\label{table:T_0}.
\end{table}

\begin{figure}
\includegraphics[width=4in]{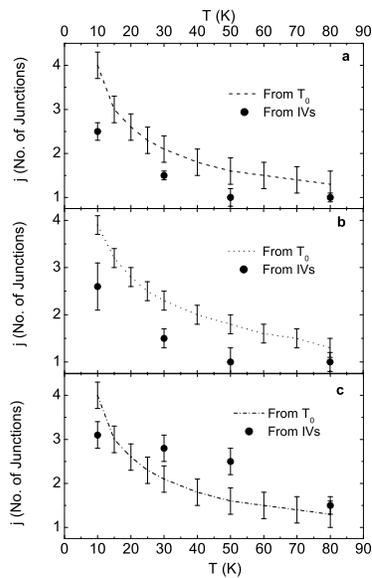}

\caption{Typical cotunnel distance as a function of temperature for trilayer (a) and tetralayer (b) arrays measured in perpendicular direction, and for a monolayer (c) measured in-plane.  The distance is given as the number $j$ of junctions along a chain of particles involved in the cooperative cotunnel process.  The plots compare the dependence extracted from the zero-bias conductance (dotted lines) with the values for $j$ obtained directly from the $I-V$ power-law exponents in the small bias regime.\label{fig:comparison}}
\end{figure}

Using $r/d = (\frac {E_C}{k_B T}\,
 \frac{\xi}{a})^{1/2}$  together with Eq.~(\ref{eq:T0}) we can estimate  the typical cotunnel distances from the experimental values for $T_0$. In Fig.~\ref{fig:comparison} we plot the result in terms of the number of junctions involved,  $j= r^*/d = (\frac {T_0}{C T})^{1/2}\,
 \frac{\xi a}{d^2}$ (dotted lines), where $C\approx 2.8$.  In the same figure, this inverse-square-root temperature dependence obtained from the zero-bias conductance is compared to the $j$-values extracted directly from the curvature of the $I-V$ characteristics at finite bias. The two sets of data track each other well, differing by less than one junction over most of the measured temperature range. For example, at 10K, the distances determined based on $T_0$ are 4 junctions for both tri- and tetra-layers, while the power-law exponents give 2.5 and 3 junctions, respectively.

At sufficiently high temperatures, the cotunnel distance approaches a single junction.  At this point $g_0(T)$ crosses over to simple activated, Arrhenius behavior, with $g_0(T) = g \exp [-\frac {U}{k_B T}]$ (dotted lines in Fig.~\ref{fig:g0T} and plot in inset).  For vertical transport in both tri- and tetra-layers this occurred above roughly 60K and led to $U/k_B \sim$ 193K and 228K, but in the bilayer was not observed over the full measured range up to 125K.  For comparison, in-plane transport data for the monolayer shown in Fig.~\ref{fig:g0T} gives a cross-over temperature $\sim70K$ and $U/k_B \sim 300$K.  Similar cross-overs at temperatures around 70-100K were observed in our previous previous data for the in-plane zero-bias conductance of bi-, tri, and tetralayers.~\cite{Tran}  From the expression~\cite{Tran, Parthasarathy2,Elteto} $U \sim 0.2NE_C$ that relates the measured high-temperature, Arrhenius-type activation energy $U$ to the charging energy we find $E_C/k_B$ values of 965K, 1140K and 1500K for the tri-, tetra-, and monolayer  in Fig.~\ref{fig:g0T}, respectively. This is in reasonable agreement with the value of  1600K we estimated above using  $E_C = \frac{e^2}{4\pi \epsilon \epsilon_0 a}$, especially given the experimental uncertainty in determining $U$ over the available limited temperature range.

\subsection{Magneto-transport}

In doped semiconductors, variable-range hopping at low bias voltage results in a change of conductance in the presence of a magnetic field. Large magnetic fields squeeze the exponentially-decaying wavefunctions in the transverse direction, i.e. if the applied magnetic field is along the out-of-plane or z-direction, the wavefunctions are squeezed in the x- and y-directions. The decrease in the wavefunction overlap gives rise to a decreased tunneling amplitude, and this in turn produces a decrease in the variable-range-hopping conductance~\cite{Shklovskii}. This phenomenon has been observed in both n-type~\cite{Halbo} and p-type semiconductors~\cite{Lee}. Recent magneto-transport data on n-type CdSe semiconductor nanoparticle films has also been interpreted within the VRH picture~\cite{Yu2}.  By investigating the magnetic field dependence of our metal nanoparticle arrays, we therefore can provide a further test for the transport mechanism: VRH is expected to produce a positive magnetoresistance, while inelastic cotunneling should not be affected even by large magnetic fields since the hops of all participating electrons involve only nearest neighbors.

\begin{figure}
\includegraphics [width = 4in]{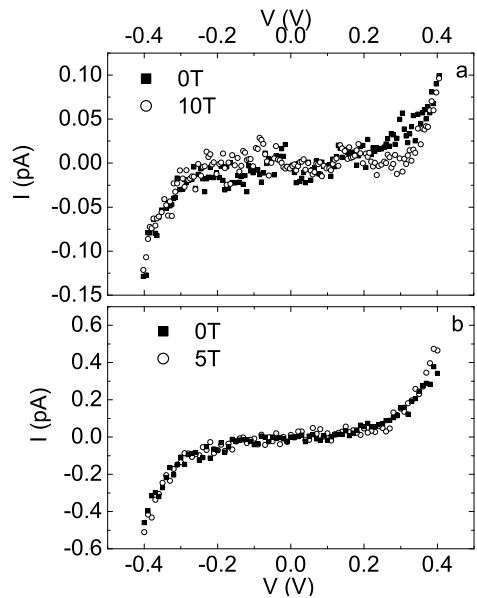}
\caption{Low-bias magnetotransport data for a monolayer measured in-plane at 400mK (a) and for a bilayer measured in perpendicular direction at 5K (b). In both cases the field was applied perpendicular to the sample plane, i.e., perpendicular to the current flow in (a) and parallel to it in (b).\label{fig:magneto}}
\end{figure}

Figure~\ref{fig:magneto} shows that the presence of a magnetic field perpendicular to the sample plane has no clear measurable effect within our experimental resolution.  These data were taken for in-plane transport through a monolayer and for vertical transport through a bilayer. The same behavior was observed when magnetic fields were applied along the direction of the sample plane. This magnetic-field-independent of the $I-V$ characteristics is to be compared to the large positive magnetoresistance ($\sim 150 \%$ at 10T and $T=0.3$K) reported for CdSe films.~\cite{Yu2} The distance over which an applied magnetic field $H$ affects charge motion is set by the magnetic length $L=(\Phi_0/H)^{\frac{1}{2}}$, where $\Phi_0 = \frac{h}{2e}$ is the flux quantum. Since $L\ge14$nm at fields $H\le10$T the absence of magnetoresistance in the weakly coupled metal nanoparticle arrays discussed here implies that the motion of individual charges occurs over distances less than $L$, in line with inelastic cotunneling. At sufficiently low temperatures  where a cross-over to elastic cotunneling takes place, magnetoresistance should reappear. In our samples, this cross-over is expected below ~1K, and it may be possible that this produced the slight differences between the traces with and without magnetic field in Fig.~\ref{fig:magneto}a. However, we believe that the data are too close to the experimental noise floor.

\section{Conclusion}
\label{concl}

The results presented here demonstrate that the electronic transport of weakly-coupled metal nanoparticle arrays is well described by a picture based on sequential tunneling in the high-bias regime above a global Coulomb blockade threshold, and  by inelastic cotunneling in the low-bias regime below that threshold.  In particular, multiple inelastic cotunneling provides an explanation for both the exponential, inverse-square-root temperature dependence of the conductance, $g_0(T)$, in the limit of zero bias and the non-linear,  power law $I-V$ characteristics observed inside the Coulomb-blockade regime at finite bias.  While the zero-bias temperature dependence of $g_0(T)$ mimics that known from single-charge, variable-range hopping transport in doped semiconductors, cotunneling events are cooperative, multi-electron processes.  In close-packed metal particle arrays individual electrons are unlikely to move further than to their nearest neighbor, at least at temperatures above ~1K, in contrast to hopping between sparse doping sites.  Comparison of transport measurements taken in-plane and perpendicular to the arrays (Figs.~\ref{fig:IVs_verticaltransport} and~\ref{fig:IVs_inplane}) indicate that typical cotunnel events in Au nanoparticle arrays involve up to 4 electrons (and thus reach net distances corresponding to 4 tunnel junctions in a row)  at 10K (Fig.~\ref{fig:comparison}). The absence of clean evidence for magnetoresistive effects up to the highest applied magnetic fields (10T) further supports the picture of inelastic cotunneling as the mechanism for charge transport inside the Coulomb blockade regime.

\section{Acknowledgments}
We thank Virginia Estevez, Klara Elteto Mueggenburg, and Andreas Glatz for insightful discussions, and Qiti Guo, Robert Josephs, Daniel Silevitch and Carlos Ancona for their help at various stages of the experiments. This work was supported by the UC-ANL Consortium for Nanoscience Research, by the NSF MRSEC program under
DMR-0213745, and by NSF grant DMR-0751473. X.-M. L. acknowledges support from DOE W-31-109-ENG-38.

\end{document}